# Robust Image Watermarking in Wavelet Domain using GBT-DWT-SVD and Whale Optimization Algorithm

**Shiva Sattarpoor[1], Hamid Barati[2]**

[1]Computer Faculty, Islamic Azad University, Dezfoul Branch, Dezfoul, Khuzestan, Iran, E-mail: sattarpour@iaud.ac.ir
[2]Computer Faculty, Islamic Azad University, Dezfoul Branch, Dezfoul, Khuzestan, Iran, Email: hbarati@iaud.ac.ir

**Abstract** As digital content can be copied easily, Copyright infringement has become a concern nowadays. Providing a solution to prevent the abuse of such contents is very necessary. One of the most common methods to solve this problem is watermarking. In this method, a logo belongs to the owner of the media is embedded in the media. So, they can prove the originality or ownership of the media content. Images are one of the most important digital media. Therefore, in this study, a method for digital image watermarking is proposed. The proposed method is based on Graph-based Transform (GBT), Singular Value Decomposition (SVD), and Discrete Wavelet Transform (DWT) which uses a Whale Optimization Algorithm (WOA) to find the best value for the embedding coefficient in the images as well as optimal blocks. The image is first transformed to a transform domain using the DWT and GBT, and then the watermark logo embedded onto the singular values of the cover image. The objective function defined for this task is based on the three parameters PSNR and NC, in the presence of image attacks. The results of the proposed algorithm on some known images show a high performance of this method compared to other similar methods.

**Keywords:** Image Watermarking, Whale Optimization Algorithm (WOA), Wavelet Domain, Graph-based Transform (GBT), Singular Value Decomposition (SVD), Discrete Wavelet Transform (DWT).

## 1 Introduction

The Internet, along with the relatively inexpensive access to digital content, has provided an environment where digital content can be easily reproduced and distributed without any loss of quality. This is a great concern to the digital publishing industry (image, music, and video) because in the past there was no technology that could safeguard the digital and material rights of digital media and prevent unauthorized copying. Although the watermarking technique can be used to prevent unauthorized access to digital content. It is clear that watermarking has its own limitations to protect the copyright. Watermarking could not prevent the copying content. But it can prove the ownership or originality of digital media. Obviously, other technologies are needed to help prove the rights of the owner, track how content is used, guarantee authorized access, and prevent unauthorized duplication of digital content [1], [2]. This need grabbed the attention of the research community and led to the birth of digital watermarking. The basic idea of digital watermarking is to create some extra data, including some information about what digital content is going to be protected, e.g. a logo of the owner. Extra data which is called watermark, is embedded in the digital content that is to be protected. In most applications watermark must be invisibly embedded in the host media. It must also be strong enough to withstand the most common signal distortions [3].

The most common algorithm used in the transform domain is a discrete wavelet transform (DWT) and single value decomposition (SVD). The main idea is that to transform the host image to another domain (wavelet or frequency domain) and then embed the watermark bits into the transformed form of the host image. It has been shown that the transformed domain using DWT or SVD is more robust against noise, compression, and other common attacks [5] [6].

In this paper, we use GBT, DWT and SVD to robust the watermarking against common attacks. Also, we use a meta-heuristic optimization algorithm, WOA [7], to optimize the parameters of the watermarking method in order to provide the desire PSNR and robustness after the embedding process.

The rest of the paper is organized as follow:

In section 2, some related studies are introduced. In section 3, important concepts such as GBT, DWT, SVD, and WOA are explained. In section 4, the proposed watermarking algorithm is presented. In section 5, experimental results are provided and finally, in section 6, we wrap up the paper with a conclusion.

## 2 Related Works

One of the common methods used in watermarking is to transform the host image into a frequency domain and then embedding the information.

Song and Qiaolun [8], have presented an algorithm for digital image watermarking based on a linear regression algorithm and a wavelet lifting scheme. They have shown that this method is robust against noise and crop attack. But they did not report the results in the presence of other attacks.

In another study [9], Sahraee and Ghofrani presented a method of watermarking using quantization of the distance between DWT coefficients. The method is blind which means that the cover image is not needed to extract the watermark logo again. In this method, first, the blocks of the cover image are transformed using DWT and three coefficients above each block are selected. After this, the selected coefficients are quantized and then the watermark image is embedded.

In [10], a blind watermarking algorithm for images is presented. This method uses the Fractional Fourier Transform (FRFT). In this study, the image is divided into non-overlapping blocks and applied an FRFT to each block. Then the binary value of the pixels is embedded in the least valuable bit of the block. The results show the poor performance of this algorithm against various attacks.

Kamble et al. [11] presented a method for digital image watermarking based on DWT-SVD using Arnold mapping. In this paper, the SVD method was used for watermarking, on the other hand, Arnold mapping was used to enhance the security of the image. One of the advantages of the method is the robustness of the method against a wide range of attacks. The results show that the method proposed in this study performs better than the SVD-based watermarking method which did not use Arnold mapping. Watermarked images also have a high quality.

Musrrat et al. [12], presented a method for wavelet watermarking based on single value decomposition (SVD) and artificial bee algorithm. The DWT for the cover image is first applied. The low-frequency sub-band is then divided into non-overlapping blocks. The blocks that are suitable for embedding the watermark are selected, the SVD coefficients of these blocks are calculated, which of the SVD coefficients can be changed to give the least change and maximum robustness in the image, by selecting the appropriate threshold value. The artificial bee optimization algorithm was used to select the optimal threshold value. The results in this paper illustrate the high robustness of this method against various attacks [12].

Musrrat and Chang [13], introduced an evolutionary algorithm, called Cuckoo Search Optimization (CSO) for finding the optimal scaling factors to robust the image watermarking. A two-level DWT on the host image is used for watermarking and CSO used to optimize scaling factors.

Saxena et al [14], proposed a watermarking method based on Discrete Fractional Fourier Transform (DFrFT) domain. In this method, the host image has transformed by DFrFT and then a binary watermark embedded into it. The adaptive Thresholding method has used for watermark extraction.

A combination of Non-Subsampled Contour-let Transform (NSCT), Discrete Cosine Transform (DCT), and Singular Value Decomposition (SVD) have introduced by Singh et al [15] for medical image watermarking. The host image in this method is a medical image and the watermark is Electronic Patient Record (EPR).

Sivaprakash et al [16], used DCT-SVD for medical image watermarking. Also, they used the Cuckoo Search Optimization (CSO) algorithm based on Levy Flight to optimize the parameters of embedding.

Zhang and Wei [17], proposed an image watermarking method based on a combination of QR decomposition and SVD which has a lower time complexity than SVD. For the robustness of watermarking they have used the gyrator transform.

Farzaneh and Mahdian [18], the combination of graph-based transform (GBT) and singular value decomposition (SVD) is used for audio watermarking and the results show that the use of GBT is more robust against various attacks, compared to other similar methods such as discrete cosine transform (DCT). They have previously showed that GBT is also better than DCT in audio compression [19]. The most important strength of this paper is that the use of GBT is introduced for the first time in the field of audio watermarking and the results show that this method can replace the powerful discrete cosine transform (DCT). It has also established a relatively good trade-off between robustness, imperceptibility and capacity. The disadvantage of this method is its low robustness to some attacks such as cropping.

In [20], a new watermarking algorithm based on GBT has introduced for Plenoptic images. In this method, each watermark bit is embedded into one micro image in transform domain. The author of [20] are also showed that GBT is more efficient than DCT in case of robustness.

Reference [21] introduces a hybrid robust watermarking method in transform domain for digital images. This method is based on discrete cosine transform (DCT), discrete wavelet transform (DWT) and singular value decomposition (SVD). One of the advantages of this method is that it is robust against rotate attack. One of the disadvantages of this method is its relatively high watermark effect on the host image. So that compared to other methods, it reduces the quality of the host image more.

In [22], a discrete cosine transformation (DCT) method is used to image watermarking. The novelty of this article is in dynamizing the intensity of embedding. Usually in watermarking methods, watermark information is added to the host image with a fixed coefficient. But in this method, an optimal coefficient is obtained for each block. The results show that this method is completely imperceptible (invisible) and has good resistance to many attacks, which is one of the advantages of the method. One of the disadvantages of this method is its low robustness against cropping attack.

In [23], a method based on Auto-Encoder neural networks is used for image watermarking. The neural network proposed in this method consists of three subnets: the encoder subnet, the decoder subnet, and the detector subnet. In the initial phase, the encoder and decoder subnets learn how to embed a bit of a watermark image into the host image. At the same time, the detector subnet learns how to retrieve 1 embedded bit. The results show that this method is highly robust against geometric attacks and noise and is also completely invisible. But it has relatively lower robustness to attacks such as rotation.

## 3 Preliminaries

As we have used Singular Value Decomposition, Discrete Wavelet Transform, and Whale Optimization Algorithm in the proposed method, in this section these three concepts have been explained.

### 3.1 Singular Value Decomposition (SVD)

Singular Value Decomposition (SVD) is a factorization of a real or complex matrix that has useful applications in signal processing and statistics. From the point of view of linear algebra, we can consider a discrete image as an array of non-negative scalar inputs that is considered as a matrix. SVD represents an image with size of $n \times n$ as $A = USV^T$. Where $U$ and $V$ are orthogonal matrices, and $S = diag(\lambda_i)$ is a diagonal matrix of eigenvalues $\lambda$ and $i = 1, 2, ..., n$, arranged in descending order. The columns of matrix $U$ are the left Eigen vectors and the columns of matrix $V$ are the right Eigen vectors. Here is an example of how SVD works:

$$A = USV^T = \begin{bmatrix} u_{11} & \cdots & u_{1n} \\ \cdot & & \cdot \\ \cdot & & \cdot \\ \cdot & & \cdot \\ u_{n1} & \cdots & u_{nn} \end{bmatrix} \begin{bmatrix} \delta_{11} & \cdots & 0 \\ \cdot & & \cdot \\ \cdot & & \cdot \\ \cdot & & \cdot \\ 0 & \cdots & \delta_{nn} \end{bmatrix} \begin{bmatrix} v_{11} & \cdots & v_{1n} \\ \cdot & & \cdot \\ \cdot & & \cdot \\ \cdot & & \cdot \\ v_{n1} & \cdots & v_{nn} \end{bmatrix}^T \quad (1)$$

Where U and V are orthogonal matrices, i.e.:

$$U \times U^T = I \quad (2)$$

SVD has particular features that make it useful for image processing, including the following:

• It has a good stability of the single S-value matrix: No significant change in the values of this matrix occurs during perturbation.
• Single values of an image determine its algebraic properties and reflect the brightness of the image. Whereas U and V represent the geometrical and spatial properties of an image.
• Single values are in descending order and many of them have small values compared to the first single. Ignoring these values has a small effect in image reconstruction.

The basic idea of SVD-based watermarking is to find the SVD of the host image or the SVD of each block of host image, and then change the eigenvalues in order to embed a watermark image in the host image. Given this, it can be said that in SVD-based watermarking methods, the eigenvalues chosen to embed, must be robust against changes and have the least impact on stability. As mentioned earlier, the S-matrix represents the eigenvalues of the image in descending order, small values of the S-matrix have less impact on the image, thus embedding the information in these coefficients does not enhance the robustness. In addition, we need to set an appropriate threshold value here to know how much change in eigenvalue must applied to make a trade-off between watermarked image quality and robustness of watermarking against various attacks.

## 3.2 Discrete Wavelet Transform (DWT)

Discrete Wavelet Transform (DWT) is a transform that captures both frequency and location information. Other transformations such as Fourier Transform (FT) just capture frequency information. So DWT has an advantage over FT. The DWT has a wide range of applications in science, engineering, mathematics, and computer science. But it mostly used in signal coding (image, audio or video compression).

The DWT of a signal $x$ is calculated by passing it through a filter bank. First the samples are passed through a low pass filter with impulse response $g$ resulting in a convolution of the two:

$$y[n] = (x * g)[n] = \sum_{k=-\infty}^{\infty} x[k]g[n-k] \qquad (3)$$

The signal is simultaneously decomposed using a high-pass filter $h$. The result provides the detail coefficients (from high-pass filter) and approximation coefficients (from low-pass filter). Since half the frequencies of the signal have now been removed, half the samples can be discarded according to Nyquist's rule. So, we have:

$$y_{low}[n] = \sum_{k=-\infty}^{\infty} x[k]g[2n-k] \qquad (4)$$

$$y_{high}[n] = \sum_{k=-\infty}^{\infty} x[k]h[2n-k] \qquad (5)$$

And then we have a subsampling by 2:

$$y_{low} = (x * g) \downarrow 2 \qquad (6)$$

$$y_{high} = (x * h) \downarrow 2 \tag{7}$$

In image processing we need to calculate the frequency horizontally and vertically both. So, in one-level DWT we will have 4 different representation of the input image. In watermarking methods that use DWT, usually the low-pass representation is selected for embedding.

### 3.3 Graph-based Transform (GBT)

All graph-based transform steps can be summarized as follows:
Having $V$ and $E$ as vertices and edges of a graph, and $s$ as a block of an image, we can define the graph $G = \{V, E, s\}$ with an adjacency matrix as following:

$$A_{ij} = \begin{cases} a_{ij}, & if\ (i.j) \in E \\ 0, & otherwise \end{cases} \tag{8}$$

Where $a_{ij}$ is the weight of the edge between vertices $i$ and $j$ in $G$.
The degree $D \in \mathbb{R}^{N \times N}$ matrix can defined as follows, which is a diagonal matrix:

$$D_{ij} = \begin{cases} \sum a_{ij}, & if\ i = j \\ 0, & otherwise \end{cases} \tag{9}$$

Using $A$ and $D$, the Graph-Laplacian Matrix can be determined as,
$$L = D - A \tag{10}$$
L is a real symmetric matrix. So, based on the spectral theory, the eigenvalue decomposition (EVD) of L have a set of real non-negative eigenvalues, and:

$$L = V \Lambda V^T \tag{11}$$

Where $\Lambda = \{\lambda_1, \ldots, \lambda_N\}$, is the set of eigenvalues, and $V = \{v_1, \ldots, v_N\}$, is the set of corresponding independent (hence, orthogonal) eigenvectors.
Using these orthogonal eigenvectors we can de-correlate the input image block, $s$, as follows:

$$c = V^T s \tag{12}$$

So, $c$ is a matrix of the GBT coefficients which has a same size as $s$.
As can be seen, the GBT is defined for one-dimensional signals, while the inputs in this study are two-dimensional image blocks. Hence, we need to apply GBT to both dimensions.
If we call the input image block, $s$, and consider the function of GBT equivalent to the equation (5) that is applied to $s$, we can use GBT in two dimensions as follows:

$$C_1 = GBT(c) \tag{13}$$
$$C_{2D} = GBT(C_1^T) \tag{14}$$

That $C_1^T$ is the transpose of a first GBT result on the input image block. Using the equation (13), the GBT is first applied to each of the rows of the image matrix, and by applying the equation (14) the GBT is applied to the columns.

## 3.4 Whale Optimization Algorithm (WOA)

Hedgehog whales can identify and surround hunting sites. Since the optimal design location in the search space is not known by analogy, the WOA algorithm assumes that the best candidate solution at the moment is target hunting or near optimal. After the best search factor has been identified, other search agents try to update their location to the best search factor. This behavior is expressed by the following equations:

$$\vec{D} = |\vec{C}.\vec{X}^*(t) - \vec{X}(t)| \tag{15}$$

$$\vec{X}(t+1) = \vec{X}^*(t) - \vec{A}.\vec{D} \tag{16}$$

Where $t$ denotes the current iteration, $A$ and $C$ are the coefficient vectors, $X^*$ is the best solution at present and $X$ is the location vector. It should be noted that if there is a better solution, $X^*$ should be updated in each iteration. The vectors $A$ and $C$ are calculated as follows:

$$\vec{A} = 2\vec{a}.\vec{r} - \vec{a} \tag{17}$$

$$\vec{C} = 2.\vec{r} \tag{18}$$

Where $a$ decreases linearly from 2 to 0 during the iterations (in both exploration and extraction phases) and $r$ is the random vector at 0 to 1.

Algorithm 1 shows the pseudo-code of WOA. For more information about WOA please refer to [7].

---

**Algorithm 1** Whale Optimization

Initialize the whales population $X_i(i=1,2,3,...,n)$
Initialize $a$, $A$, and $C$
Calculate the fitness of each search agent
$X^*$=the best search agent
**Procedure** WOA(*population, a, A, C, MaxIter,…*)
   $t=1$
  **while** $t \leq MaxIter$ **do**
     **for** each search agent **do**
        **if** $|A| \leq 1$ **then**
           Update the position of the current search agent by eq. 8 and 9
        **end if**
     **end for**
     Update $a, A, C$
     Update $X^*$ if there is a better solution
     $t=t+1$
  **end while**
  **return** $X^*$
**end procedure**

## 4 Proposed Watermarking Algorithm

This section describes the proposed method for embedding the Watermark image in the host image and extracting it from Watermarked image.

### 4.1 Embedding

The host image is considered to be an RGB image, and the watermark image is a binary image. First of all, the blue channel of the RGB image is selected for watermarking, and then it divides into 8×8 blocks. Each block can take one bit of watermark image. For each selected block, a 2D DWT using the Haar wavelet is applied on the block. Then we choose the LL (low-low) component and apply GBT and then an SVD to it. We just use the biggest singular value (first element of matrix S) for embedding by means of this equation:

$$S_w = \begin{cases} S_h + \alpha, & if\ watermark\ bit = 1 \\ S_h - \alpha, & if\ watermark\ bit = 0 \end{cases} \quad (19)$$

Where $S_w$, is the biggest singular value of the watermarked image, $S_h$ is the biggest singular value of the host image, and α is a constant value that should be optimized by WOA. We also need to store the $S_h$ value in each block for the extraction purposes.

Now we can use $S_w$, inverse of SVD, and inverse of GBT to make the LL component again. Then we apply the inverse of DWT to make the watermarked block. This process is repeated for all selected blocks and all watermark bits. After that, we create the watermarked blue channel using the watermarked blocks and then we can provide the watermarked RGB image. Fig.1 demonstrates the embedding flowchart. In the proposed method, we use the optimal key for efficient block selection, and we use the optimal α for efficient embedding. These optimal values are calculated by the Whale Optimization Algorithm.

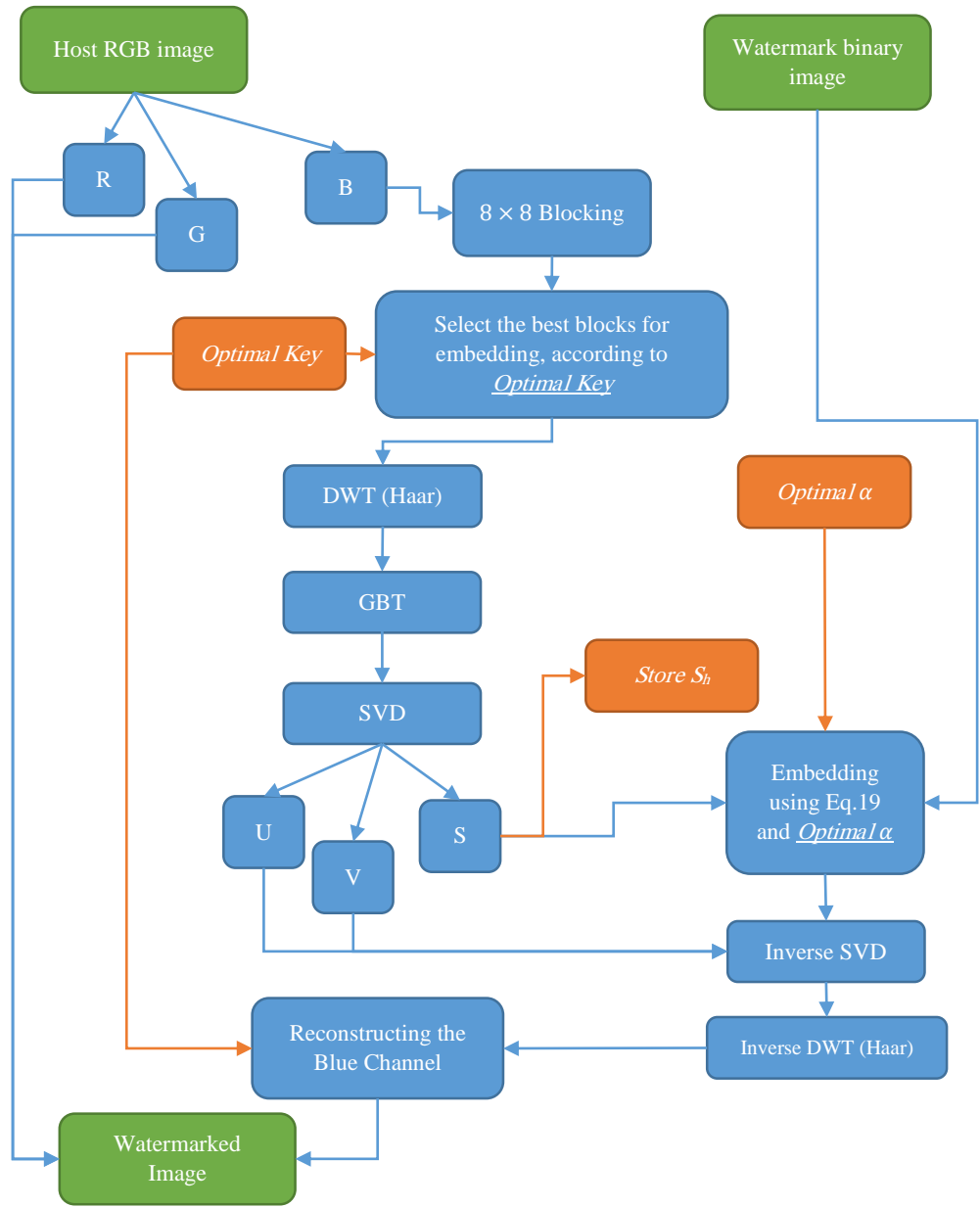

**Fig. 1** Embedding Flowchart

## 4.2 How to use the whale optimization algorithm in watermarking

In the previous section, α is referred to be the embedding value. The value firstly, should not affect the quality of the host image, and secondly, it should not disappear in the presence of various attacks such as noise and compression. Also, we have several choices to select the blocks. It is clear that all blocks are not appropriate for embedding. So, we need to find the optimal blocks.

To find the best blocks and also best α values, we have used WOA. WOA should find the best blocks and best α values for each block. Best blocks are considered as optimal key. It shows that each watermark bit should be embedded in which block of the host image. Also, we need to find the optimal α values for each block. So, every search agent in WOA is a vector with a length of 2M where M is the number of watermark bits.

The following equation describes the objective function defined to solve the problem.

$$\underset{(\alpha,key)}{minimize} f = 10 \times |PSNR - PSNR_{target}| + \left(1 - \frac{1}{N}\sum_{i=1}^{N} NC_i\right) \quad (20)$$

Where $N$ is the number of attacks we encounter, $NC_i$ is normalized correlation between watermark image and extracted watermark after attack $i$. The attacks defined for this research are average filter, median filter, salt and pepper noise, Gaussian noise, re-scaling and MPEG compression.

$$NC(W, W^*) = \frac{\sum_{i-1}^{n}\sum_{j-1}^{n}\overline{W_{i,j} \oplus W^*_{i,j}}}{n \times n} \quad (21)$$

Where $W$ is the Watermark image, and $W^*$ is the extracted image.
The PSNR considers the watermarked image quality. $PSNR_{Target}$ is the desire PSNR.

$$PSNR(I, I') = 10 Log_{10}\left(\frac{(I_{Max})^2}{\frac{1}{n \times n}\sum_{i-1}^{n}\sum_{j-i}^{n}(I_{i,j} - I'_{i,j})^2}\right) \quad (22)$$

Where $I$ and $I'$ are original and watermarked images. $I_{max}$ is the maximum possible intensity value for Image $I$.

### 4.3 Extracting

The proposed method is semi-blind. Because we need $S_h$ values) a part of host image for the extraction. All the steps in embedding process i.e. blocking, DWT, GBT, and SVD is applied on watermarked image then we extract the watermark bit using following equation:

$$Watermark\ bit = \begin{cases} 1 &, if\ S_w > S_h \\ 0 &, if\ S_w \leq S_h \end{cases} \quad (23)$$

Where $S_w$, is the biggest singular value of the watermarked image, and $S_h$ is the stored biggest singular value of the host image. Fig.2 demonstrates the extracting flowchart.

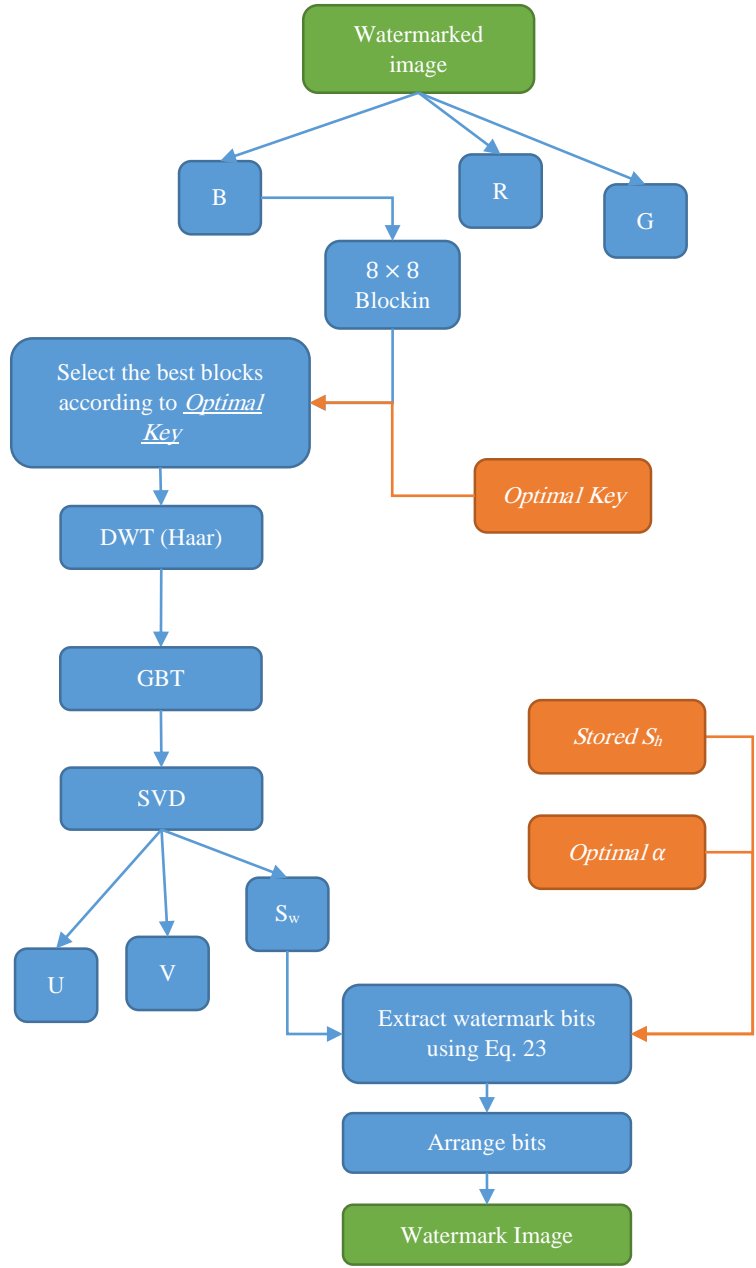

**Fig. 2** Extracting Flowchart

## 5 Experimental Results

We have used MATLAB R2018b for the implementation of the proposed watermarking algorithm on Windows10 operating system. Table 1 shows the parameters settings.

Table 1 Parameters setting for the proposed method

| PARAMETER | VALUE |
| --- | --- |
| Number of Search Agents in WOA | 100 |
| Maximum Iteration in WOA | 1000 |
| $PSNR_{Target}$ | 45 dB |
| Number of Attacks | 6 |
| Host image size | $256 \times 256$ or $1024 \times 1024$ |
| Watermark size | $20 \times 20$ or $80 \times 80$ |
| Block size | $8 \times 8$ |

For the optimization step, we have used a constant watermark image (Panda logo), and for other experiments, we have used random binary watermarks. Also we have used five popular images for the hosts. I.e. Baboon, Barbara, Boats, Peppers, and Lena. Figure 3 and Figure 4 show the host and watermark images, respectively.

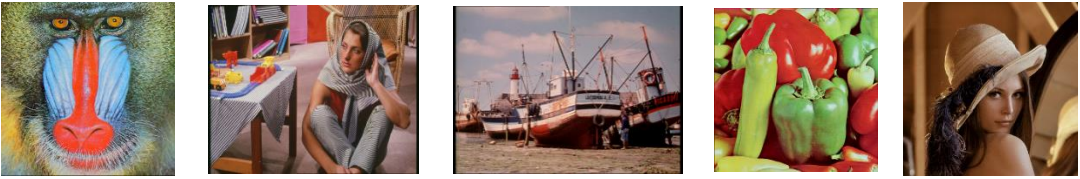

Fig. 3. Host images

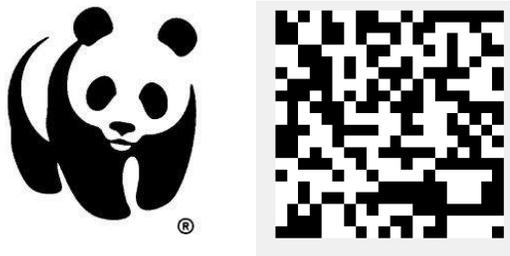

Fig. 4. Watermark Images

Table2 has summarized the results of invisibility. In Table2 Peak Signal to Noise Ratio (PSNR) and Structural Similarity Index Measure (SSIM) are between watermarked image and the host image.
Table3 reports the Bit Error Rates (BER) between the original watermark image and the extracted watermark image.
Table4 compares the proposed method with some other studies on Barbara RGB image.

Figure 5 shows some examples of watermark extraction in presence of various attacks for Lena Host image and Panda logo watermark image, and figure 6 shows the average value of BER for all experiments on different host and watermark images.

Table 2 Invisibility results for different images

| Images | PSNR (dB) | SSIM |
|---|---|---|
| Baboon | 70.99 | 0.9947 |
| Barbara | 71.33 | 0.9909 |
| Boats | 71.38 | 0.9927 |
| Peppers | 73.24 | 0.9920 |
| Lena | 74.93 | 0.9872 |

Table 3 Bit Error Rates for different Images in presence of various attacks

| Image | No attack | Gaussian Noise | Salt & Pepper | Speckle Noise | Median Filter | Re-Scale | MPEG |
|---|---|---|---|---|---|---|---|
| Baboon | 0 | 0 | 0.0075 | 0.0425 | 0.0350 | 0.0725 | 0 |
| Barbara | 0 | 0 | 0.0075 | 0 | 0.0075 | 0 | 0 |
| Boats | 0 | 0.0225 | 0.0300 | 0 | 0.0200 | 0.0250 | 0.0250 |
| Peppers | 0 | 0.0250 | 0.0575 | 0 | 0.0250 | 0.0150 | 0.0100 |
| Lena | 0 | 0.0475 | 0.1100 | 0 | 0.0300 | 0.0550 | 0.0075 |
| Average | 0 | 0.0190 | 0.0425 | 0.0085 | 0.0235 | 0.0335 | 0.0058 |

Table 4 Proposed Method comparing to other studies on Barbara image in case of Bit Error Rate (BER)

| Attack Type | Proposed | Liu and Tan [28] | Fazli and Moeini [29] | Xiao et al. [30] |
|---|---|---|---|---|
| Gaussian Noise | **0** | 0.176 | 0 | 0.0014 |
| Salt and Pepper Noise | 0.0075 | 0.084 | **0** | 0.0015 |
| Re-Scale | **0** | 0.087 | 0.04 | 0.0328 |
| Median Filter | **0.0075** | 0.1500 | 0.038 | 0.0280 |

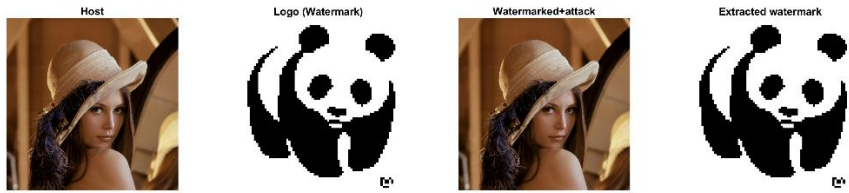

Attack 1. No Attack

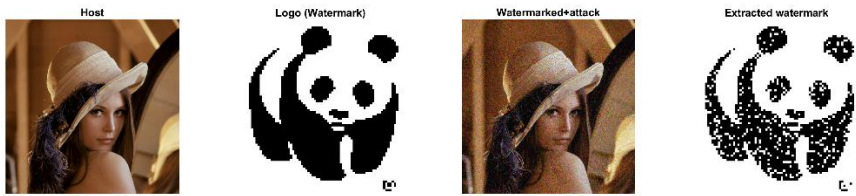

Attack 2. Gaussian Noise

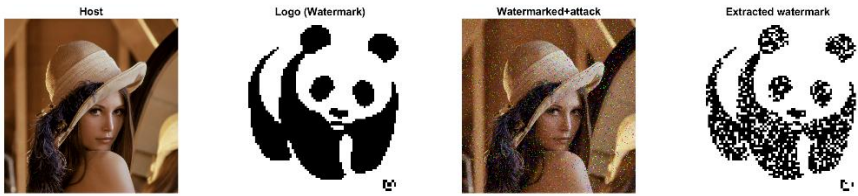

Attack 3. Salt and Pepper

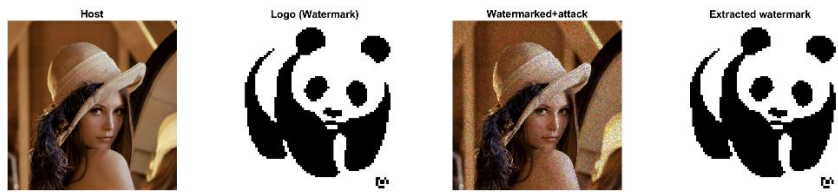

Attack 4. Speckle Noise

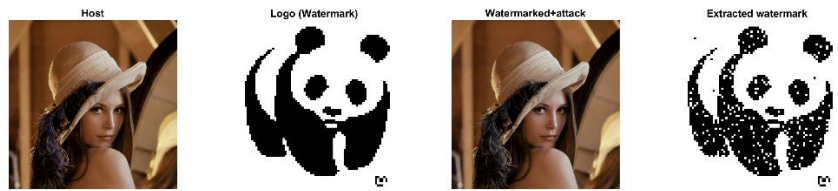

Attack 5. Median Filter

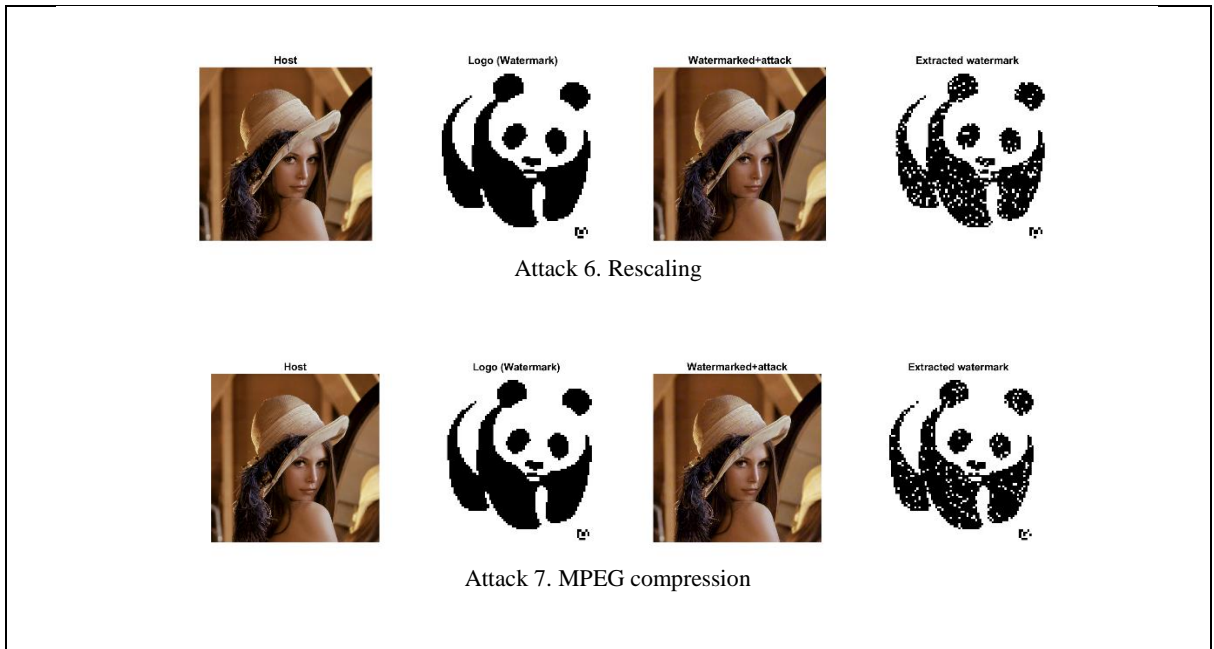

Attack 6. Rescaling

Attack 7. MPEG compression

Fig. 5. Watermark Extraction in presence of various attacks for Lena image

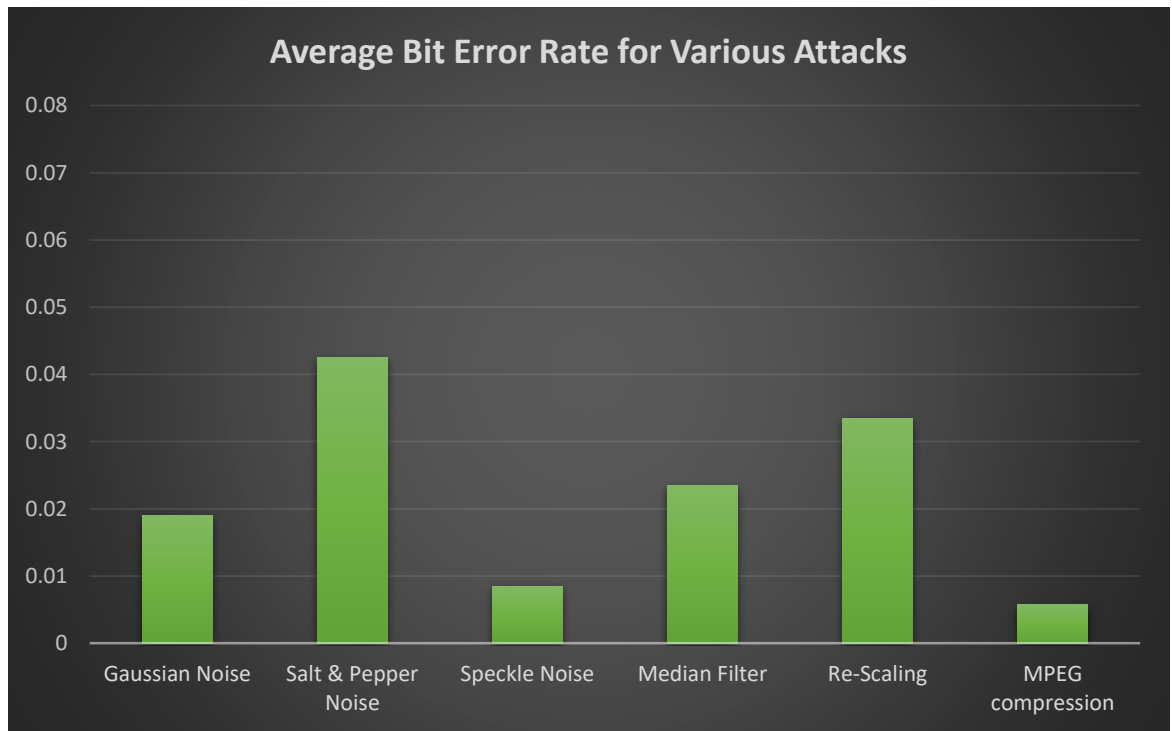

Fig. 6 Average BER of proposed method against various attacks

## 6 Conclusion

In this paper, we proposed a new image watermarking method based on discrete wavelet transform, graph-based transform, and singular value decomposition. In this method, we used the Whale Optimization Algorithm to optimize the watermarking parameters and to find the best blocks. The results show well that the proposed method, in addition to maintaining the quality of the host image and performing a completely invisible watermarking, also has a high robustness against various types of attacks such as noise, compression, rescaling and filters.